\documentclass[10pt,letterpaper]{article}

\usepackage{cogsci}

\cogscifinalcopy 

\usepackage{booktabs}
\usepackage{graphicx} 
\usepackage{pslatex}
\usepackage{apacite}
\usepackage{natbib}
\usepackage{float} 
\usepackage{algorithm}
\usepackage{algpseudocode}
\usepackage{amsmath}
\usepackage{blindtext}

\usepackage{xcolor}
\usepackage{hyperref}
\hypersetup{
    colorlinks,
    linkcolor={red!50!black},
    citecolor={blue!50!black},
    urlcolor={blue!80!black}
}
\usepackage{amsfonts}

\global\hyphenpenalty=1000

\newcommand{\fig}[1]{Fig.~\ref{#1}}

\renewcommand{\AE}{\textrm{AE}}
\newcommand{\ksamples}{20}
\newcommand{\argmax}{\textrm{argmax}}
\newcommand{\argmin}{\textrm{argmin}}
\newcommand{\SelectNode}{\textrm{SelectNode}}
\newcommand{\ExpandNode}{\textrm{ExpandNode}}
\newcommand{\Backpropagate}{\textrm{Backpropagate}}
\newcommand{\DepthCharge}{\textrm{DepthCharge}}
\newcommand{\PlacePiece}{\textrm{PlacePiece}}
\newcommand{\HasTerminated}{\textrm{HasTerminated}}
\newcommand{\LegalMoves}{\textrm{LegalMoves}}
\newcommand{\RandMove}{\textrm{RandMove}}
\newcommand{\Node}{\textrm{Node}}
\newcommand{\IsWin}{\textrm{IsWin}}
\newcommand{\IsDraw}{\textrm{IsDraw}}

\title{People use fast, goal-directed simulation to reason about novel games\thanks{Presented at CogSci 2024 as a talk.}} 

\author{{\large \bf Cedegao E. Zhang\textsuperscript{1,*}} \\
    \texttt{cedzhang@mit.edu} \\
    \And {\large \bf Katherine M. Collins\textsuperscript{2,*}} \\
    \texttt{kmc61@cam.ac.uk}
    \And {\large \bf Lionel Wong\textsuperscript{1,*}} \\
    \texttt{zyzzyva@mit.edu} \\ 
    \And {\newline \large \bf Mauricio Barba\textsuperscript{1}} \\
    \texttt{barba@mit.edu} \\
    \AND {\large \bf Adrian Weller\textsuperscript{2,3}} \\
    \texttt{aw665@cam.ac.uk}
    \And {\large \bf Joshua B. Tenenbaum\textsuperscript{1}} \\
    \texttt{jbt@mit.edu} \\
    \AND
    \textsuperscript{1}MIT BCS,
    \textsuperscript{2}University of Cambridge,
    \textsuperscript{3}The Alan Turing Institute \\
    \textsuperscript{$\ast$}These authors contributed equally to this work.
    }

\begin{document}

\maketitle

\begin{abstract}

People can evaluate features of problems and their potential solutions well before we can effectively solve them. When considering a game we have never played, for instance, we might infer whether it is likely to be challenging, fair, or fun simply from hearing the game rules, prior to deciding whether to invest time in learning the game or trying to play it well. Many studies of game play have focused on optimality and expertise, characterizing how people and computational models play based on moderate to extensive search and after playing a game dozens (if not thousands or millions) of times. Here, we study how people reason about a range of simple but novel Connect-N style board games. We ask people to judge how fair and how fun the games are from very little experience: just thinking about the game for a minute or so, before they have ever actually played with anyone else, and we propose a resource-limited model that captures their judgments using only a small number of partial game simulations and almost no look-ahead search.

\textbf{Keywords:} 
intuitive theories; game theory; reasoning; problem solving; computational modeling
\end{abstract}

\section{Introduction}

\begin{figure}[ht]
\begin{center}
\includegraphics[width=0.5\textwidth]{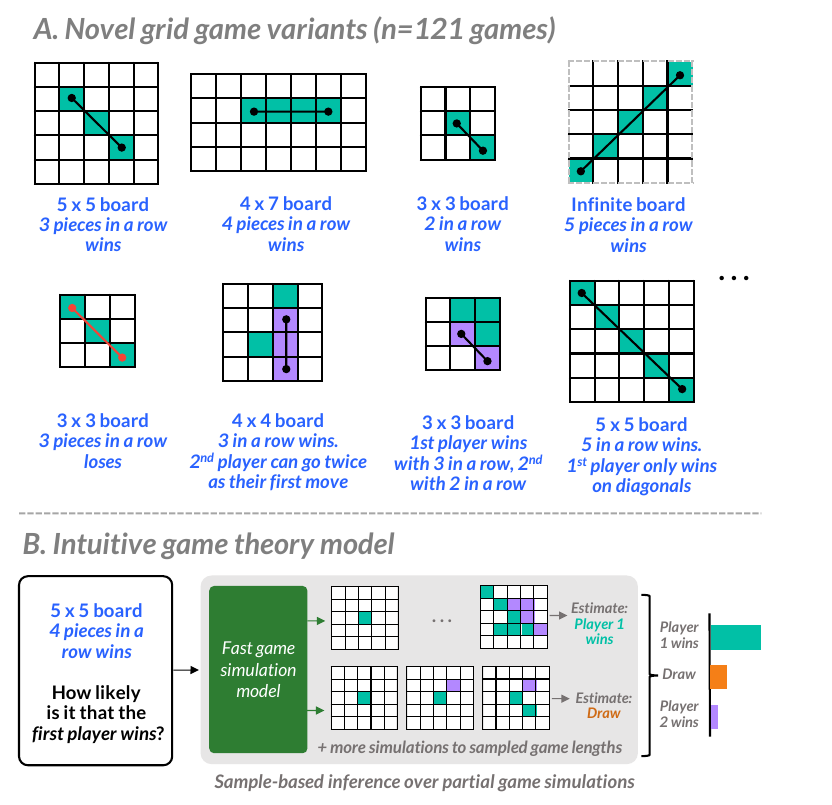}
\end{center}
\caption{(\textbf{A}) Design of 121 grid games varying the game environment, dynamics, and win conditions. (\textbf{B}) Our \textit{intuitive game theory} model simulates bounded game play under a fast but general agent model to draw probabilistic inferences about novel games.} 
\label{fig-novel-game-theory}
\end{figure}

Consider the two-player game shown in the top left of \fig{fig-novel-game-theory}, where players alternate placing their pieces in grid squares and winning means being the first to connect \textit{3 pieces in a row on a 5 by 5 board}. Though you may be familiar with related games like Tic-Tac-Toe or Connect-4, chances are you have not played this particular game before. Still, with just a moment's thought, you can likely evaluate some basic but important aspects of what it would be like to actually play this game against a reasonable opponent---how many moves would it take for the game to finish? Would you rather go first or second? Would this be fun to play with another novice? Think for a little longer, and if you hadn't already, you might realize that the game is actually \textit{very} biased in favor of the first player, and perhaps not fun at all to play for more than a few rounds. You could probably evaluate any of the games at the top of \fig{fig-novel-game-theory} in the same way, and you likely \textit{would} think about these kinds of questions before deciding to master one of these games as a hobby, or to bet money on their outcomes.

The formal and empirical study of games often focuses on \textit{optimality} and \textit{expertise}. Game theory in mathematical economics, as studied by \cite{nash1950equilibrium} for example, largely seeks to characterize equilibrium states reached under rational play. Computational game playing models, from classic AI expert systems like Deep Blue to more recent neurally guided models like AlphaGo and AlphaZero \citep{silver2016mastering,silver2017mastering_go,silver2018genearl}, aim to match or exceed the abilities of human experts, generally by evaluating millions of simulated games. A rich body of related cognitive science research documents how human experts play after extensive experience on a given game \citep{newell1972human, gobet2004moves}. 
Experts seem to learn highly specialized and game-specific representations, like memorized chess states and openings, that enable them to effectively search or approximate the results of searching much further ahead in a game than novice players \citep{chase1973perception,van2023expertise}.

But we simply \textit{aren't} experts on most problems we encounter. While we might master a few domains over a lifetime, the sheer range of environments and tasks we take on also seems to demand capacities for efficiently but intelligently \textit{evaluating} a much broader range of novel problems under limited experience. How do we predict, for instance, whether a given goal or cooperative task will be tractable, rewarding, or fair enough to invest time and resources in solving it more completely? 

The present study focuses on this higher-level capacity to flexibly evaluate novel inference and decision making problems, using a set of \textit{novel strategy games} that vary the environment, dynamics, and win conditions of more familiar grid games as a test bed for general problem evaluation. We propose a computational model (\fig{fig-novel-game-theory}b) that implements novice but goal-directed agents as \textit{fast but search-limited} planners. Our model estimates the value of intermediate game state and move pairs based on simple but general concepts about game play, such as making progress towards a sparse overall goal (like making $M$ in a row) and preventing opponent progress when in a competitive setting. We then model how people evaluate games overall by nesting this agent model within a sample-based inference procedure that estimates game outcomes from a limited number of \textit{partial game simulations}. While we focus here on games, the underlying building blocks of this ``intuitive game theory'' model are designed to capture more basic multi-agent planning and probabilistic inference components applicable to a much broader range of problems---where people are called on to make good guesses and good bets in novel situations, and to evaluate their prospects of success along with the emotional reward of engaging \citep{allen2024using}.

To test our model, we ask human subjects to evaluate \textbf{game outcomes} for our novel game stimuli, such as how likely the first player is to win or draw in any give game, and to predict \textbf{how fun a game would be} for people to play. We allow participants to use an interactive \textit{scratchpad} board in which they can place pieces during the experiment, collecting data on when and how they may choose to simulate games against imaginary opponents. 

We find that our intuitive game theory model, despite using only a single step of search in its agent model and a small number of partial game simulations to estimate game play, well predicts participants' judgments of game outcomes on novel games. We compare our model against several alternative models, which either (i) consider even more naive game agent models, or conversely (ii) expend much \textit{more} computational effort on search and game simulation to more optimally predict outcomes. We also compare against two language-model-based baselines with no explicit agent model that reason solely from background knowledge about other more familiar games.

Using our model, we also show that several game-related factors correlate with human \textit{fun} ratings on these games: a \textit{fairness} factor based on estimated entropy over game outcomes; a \textit{challengingness} factor based on the estimated advantage our model has in playing against a random agent; and a fun factor based on language model predictions. Combined into a joint regression model, these factors together can predict most of the variance in human fun judgments.

\begin{table*}[ht!]
\small
\centering
\resizebox{\textwidth}{!}{%
\begin{tabular}{@{}llr@{}}
\toprule
\textbf{Category}      & \textbf{Example(s)}                               & \textbf{Count} \\ \cmidrule{1-3}
\textbf{M in a row on square boards}      & \textit{3 pieces in a row wins on a $6 \times 6$ board; 7 pieces in a row wins on a $10 \times 10$ board}                               & 20 \\ \cmidrule{1-3}
\textbf{M in a row on rectangular boards} & \textit{3 pieces in a row wins on a $1 \times 5$ board; 6 pieces in a row wins on a $5 \times 10$ board}                              & 18 \\ \cmidrule{1-3}
\textbf{Infinite boards}                         & \textit{3 pieces in a row wins on an infinite board; 10 pieces in a row wins on an infinite board}                  & 3 \\ \cmidrule{1-3}
\textbf{M in a row loses}                       & \textit{A player loses if they make 5 pieces in a row, on a $5 \times 5$ board.}                                          & 10 \\ \cmidrule{1-3}
\textbf{No diagonal wins allowed}               & \textit{4 pieces in a row wins on a $10 \times 10$ board, but a player cannot win by making a diagonal row.}                   & 10 \\ \cmidrule{1-3}
\textbf{Only diagonal wins allowed}             & \textit{4 pieces in a row wins on a $5 \times 5$ board, but a player can only win by making a diagonal row.}                         & 10 \\ \cmidrule{1-3}
\textbf{First player moves 2 pieces}            & \textit{5 pieces in a row wins on a $10 \times 10$ board; the first player can place 2 pieces as their first move.}          & 10 \\ \cmidrule{1-3}
\textbf{Second player moves 2 pieces}           & \textit{10 pieces in a row wins on a $10 \times 10$ board; the second player can place 2 pieces as their first move.}        & 10 \\ \cmidrule{1-3}
\textbf{First player handicap (A)}              & \textit{3 pieces in a row wins on a $3 \times 3$ board, but the first player cannot win by making a diagonal row.}           & 10 \\ \cmidrule{1-3}
\textbf{First player handicap (B)}              & \textit{4 pieces in a row wins on a $7 \times 7$ board, but the first player can only win by making a diagonal row.}         & 10 \\ \cmidrule{1-3}
\textbf{Second player needs M-1 to win}         & \textit{The first player needs 4 pieces in a row, but the second player only needs 3 pieces to win on a $5 \times 5$ board.} & 10 \\ \bottomrule
\end{tabular}%
}

\caption{Example games from our stimuli, summarized by categories varying game environment, dynamics, and win conditions.}
\label{tab:stimuli}
\end{table*}

\section{Intuitive Game Theory Computational Model}
How do we quickly evaluate key aspects of a novel problem--such as whether it is potentially rewarding or hugely unfair--before investing time and effort to acquire longer-term expertise? Even for the relatively simple grid games we study here, estimating potential rewards under \textit{optimal} or expert-like game play presents several related computational challenges.

To choose any \textit{single} action, an optimal agent model should estimate and then seek to maximize downstream utility with respect to all available actions from the current game state. Many general game play models estimate utility for novel games using expensive look-ahead search, ideally to terminal reward-generating game states \citep{genesereth2014general, yannakakis2018artificial}. For instance, the class of games we consider here defines very sparse utility functions---winning or losing is only determined based on a final state in which a player or their opponent has placed $M$ pieces in a row. In cases where $M$ is large or players are playing on large boards (e.g., the goal is to make \textit{8 in a row on a 20 by 20 board}), this means searching over potentially long trajectories of game play to evaluate any single action.

To then draw probabilistic judgments about game outcomes (such as how likely a given non-deterministic player is to win on average), generic sample-based inference based on this agent model would require running multiple full game simulations, wherein \textit{each} agent action would require downstream search to these terminal states.

In contrast, our \textit{intuitive game theory} model aims to produce fast, probabilistic inferences over a broad range of novel games. This model simulates agent game play based on basic but general assumptions about goal-oriented players that can be estimated directly from intermediate states, then draws graded judgments based on \textit{partial game simulations} over a range of game depths rather than always simulating to terminal states. 

\subsection{Game specifications and game reasoning queries}
We begin by defining a general notion of an \textit{intuitive game theory} problem that encompasses the varying questions we seek to evaluate (e.g., \textit{how likely is this game to end in a draw?}) on arbitrary games. We formalize a game specification as comprising an \textit{environment definition} (e.g., the board shape), a \textit{game state transition function} that defines the game dynamics (e.g., alternating actions over two players, and terminal states based on the win or draw conditions), and \textit{game utility functions} defined for each player given game state, which in our dataset are only sparsely defined over terminal states. This formalization derives from standard multi-agent planning problems definitions used in the AI planning literature (e.g., \citet{russell2020artificial}).

A \textit{problem} is then a game specification combined with any general game reasoning query about that game.

\subsection{Estimating game outcomes by simulating goal-directed but search-limited players}

The core of our model is a simulated agent that can play from game specifications. This agent model has a general \textit{search-limited} way of assigning values over possible moves given current board states.

We design the agent model around a collection of value functions which capture general intuitions about goal-directed competitive game play, and can be quickly evaluated over any intermediate board state with at most a \textit{single} step of look-ahead search in the class of games we consider here. Together, these value functions are closely related to features in models from a long tradition of works that study game-playing in \textit{specific} ``$m$-in-a-row'' games \citep{amir2022adaptive, crowley1993flexible, van2023expertise}, which are often descriptively derived from empirical observations of game play. Here, we generalize these game-specific features into a model applicable across the entire, broad family of games we consider in this paper, and also ground each in more general intuitions about game play. The general rationales beyond the features also admit other value function formulations that could be explored in future work.

\begin{figure*}[h!]
\begin{center}
\includegraphics[width=0.95\linewidth]{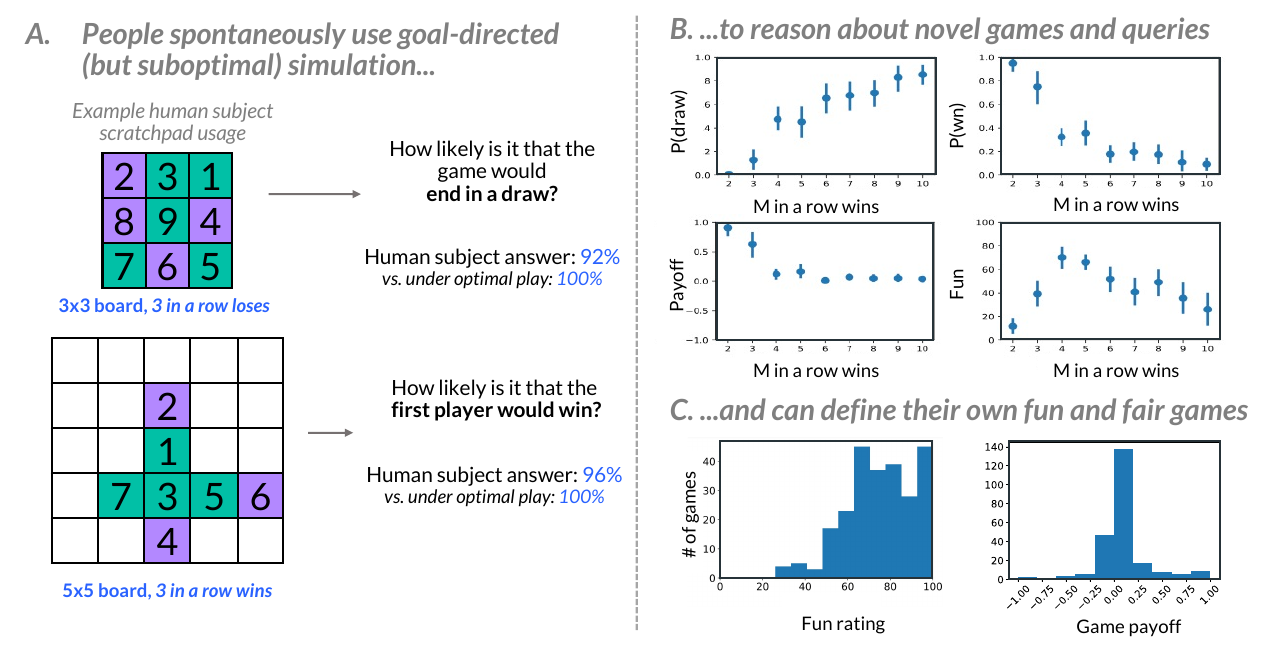}
\end{center}

\caption{(A) Qualitative analysis of voluntary participant usage of an interactive grid ``scratchpad" suggests that many participants spontaneously seem to simulate game play against imaginary opponents to reason about novel games, though these simulations are often far from optimal: the upper game could be provably drawn by first playing in the center and then mirroring opponent play; the lower game shows an inefficient win strategy. (B) Human game evaluations of game outcomes, including the probability that the game ends in a draw, that the first player wins, and the expected payoff; and a rating of how fun a game is, on the simple class of Tic-Tac-Toe extensions (winning with \textit{M in a row on an N by N board}). (C) People can generate their \textit{own} novel game variants that they generally perceive to be reasonably fun, and rate fun games as fair ones (with an expected balanced payoff of 0, rather than a biased one).} 
\label{results-hvm}
\end{figure*}

Specifically, given an intermediate board state, our player model assigns an overall value score to each possible next move (each open position on the board) based on the following functions:
\begin{enumerate}
    \item The normalized Euclidean distance of the position and the center of the board, $d \in [0, 1]$. This reflects the intuition, applicable across our family of games, that people often but not always place pieces around the center at the beginning. This function can be explained based on a more general rationale---agnostic to other aspects of the current game state, placing pieces closer to the center of the board allows that piece to participate in the \textit{most possible winning terminal states} for any \textit{$m$-in-a-row} win condition.
    
    \item The maximum number of live connected (contiguous) pieces that the position entails on any of the \textit{allowed} winning directions (horizontal, vertical, or diagonal in general, modulated with respect to any player-specific restrictions on win directions), $n_1 \in \mathbb{Z}^+$. By \textit{live} we mean connected pieces that have enough open positions on the corresponding direction such that they can be extended to form a winning \textit{$m$-in-a-row}. For example, if $m = 4$, then three connected pieces that are blocked by the opponent on both sides are not live. If $n_1$ equals the winning $m$, we add an additional 1 to $n_1$ to magnify the value of winning with respect to other states. This function captures the general intuition that players are \textit{goal-directed} and try to make \textit{progress towards winning}. That is, players derive utility from intermediate states towards the most rewarding terminal state. While there are many possible subgoals that a player could consider which might make partial progress towards terminal states (e.g., placing \textit{3 in a row} by placing two unconnected pieces and then placing one more piece in the center), we choose a simple, easy to calculate function that reflects a straightforward intuition (making $m$ in a row by previously building up to $m-1$ in a row, $m-2$ in a row, and so on via connected pieces). Variants of this function could be studied in future work, some of which could reflect more expert (rather than naive) play. For instance, we might reward positions based on the total number of live connected pieces that the position entails across all of the directions.
    
    \item Relatedly, the maximum number of live connected contiguous pieces that the position blocks the opponent from having, $n_2 \in \mathbb{Q}^+$, also considered with respect to the win-directions applicable to the opponent. This function captures the intuition that competitive players in antagonistic two-player games should also attempt to \textit{block opponent progress towards opponent goals}. Note that this function is simple, symmetric to our definition of player progress (because players in our games have related and opposing goals). It also effectively evaluates just a single step of look-ahead search relative to a opponent who is also relatively naive (that is, an opponent who at the next step would simply attempt to make progress towards their nearest victory along a simple policy). We subtract $0.5$ from $n_2$ to reflect people's tendency to weigh offense over defense \citep{crowley1993flexible}, so blocking the opponent's $\hat{n}$ in a row is not as good as making an $\hat{n}$ in a row for oneself (but is better than making an $\hat{n} - 1$ in a row). However, if $n_2$ equals the winning $m$, we do not subtract $0.5$, similarly to upweight winning. One could consider other related or more complex formulations that also capture the notion of blocking opponent progress.
\end{enumerate}

\begin{figure*}[h!]
\begin{center}
\includegraphics[width=0.95\linewidth, trim = {0 1cm 0 1cm}]{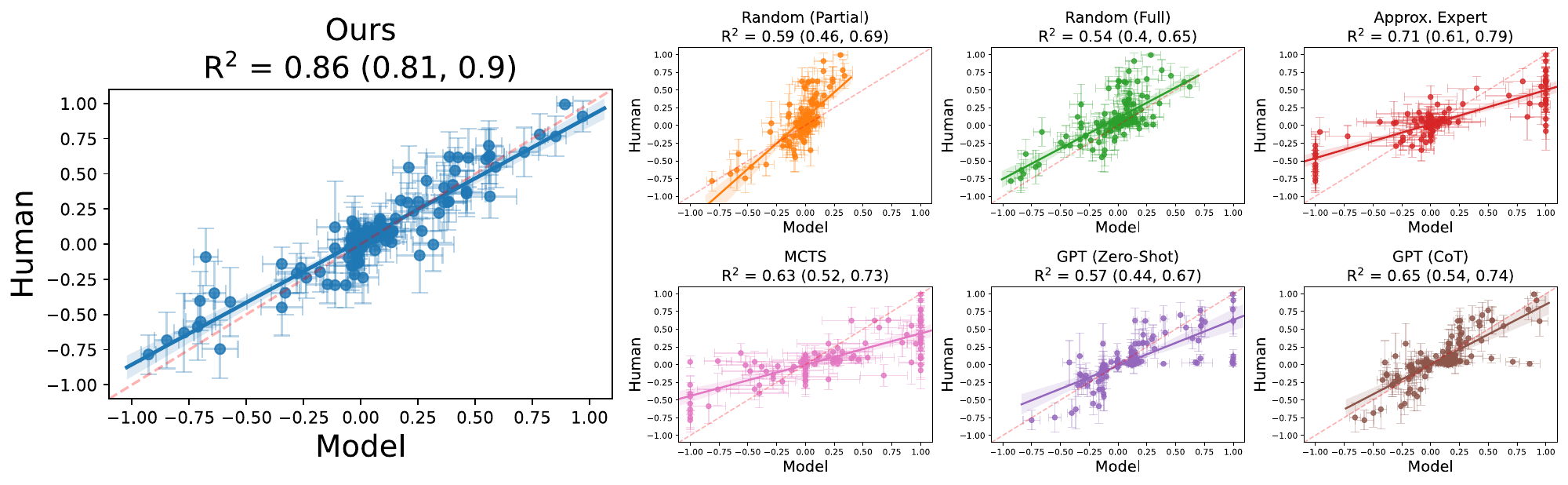}
\end{center}
\caption{Expected payoff across models against humans-predicted payoff. Each point represents the payoff for one of the $n=121$ game stimuli. Error bars in the y-axis are over individual human judgments per game and depict standard error. The $R^2$ values in the parentheses indicate 95\% CI.} 
\label{overall-payoff}
\end{figure*}

These components together reflect the intuitions that people often but not always place pieces around the center at the beginning, people naturally try to make progress towards winning, and people know they should secure winning moves and block the opponent from having those. The value of a position $p$ in board state $s$ is then given by a sum of exponential terms based on the components:
\begin{equation}
    \label{eq:subgoal}
    \begin{split}
V(p, s) = 2^{(1-d)} + 2^{n_1} + 2^{n_2.}
    \end{split}
\end{equation}

We apply a \texttt{softmax} (temperature = $1$) to the set of values to obtain a probability distribution over possible moves in a state, and the agent samples a move from this distribution.

This agent model simulates game \textit{play}---it captures intuitions about how a rational but relatively naive, resource-bounded player might choose an action from any given board state, on any of the games in our dataset. Given this model, how might someone draw graded inferences \textit{about} the game itself, like about how fair the game is or how likely a player is to reach a particular outcome? We give a straightforward ``game inference" model by nesting our agent model inside a sample-based probabilisitic inference model, which uses simulations of game play to estimate these kinds of features of a given game. 

More specifically, for any given game, our model estimates inferences about any specific game query (e.g., \textit{what is the likelihood of any specific game outcome, win, loss, or draw}; \textit{what is the likelihood that the first player will encounter any particular outcome}?) by evaluating these queries with respect to $k$ sampled simulations of two players playing the game. This is a probabilistic estimator, as our agent model is stochastic. In keeping with our focus on fast, resource-limited reasoning about games, we specify two additional parameters that can bound the resources involved in inference. First, we estimate inference queries with a \textit{small number of game simulations} (we use $k=\ksamples$ in our reported results), building upon the idea that people use a relatively small number of simulations to form beliefs and make decisions \citep{vul2014one, icard2016subjective}. Second, we allow our model to draw estimates using \textit{partial simulations}: for each simulation, we randomly sample a bound on the total number of moves we will simulate (we use a uniform distribution over moves from $1$ to the size of the game board). We estimate the game outcome of any given simulation only based on the board state reached in that bounded set of moves (if a terminal win state hasn't been reached, we simply score the game as a draw, though future work could use other estimators based on our heuristic board-state utility estimate). This bound reflects the notion that people might in fact draw estimates based on incomplete simulation of game play, a behavior qualitatively observed in the human scratchpad usage data. 

This general inference algorithm, using a small number of partial, randomly bounded simulations, themselves simulated using a search-limited agent model, can be used to estimate many different game outcome queries. We focus on a standard one, expected \textit{payoff} for the game (where we say \textit{winning} incurs a reward of $+1$, \textit{losses} incur $-1$, and draws are $0$). 

\section{Human and Model Experiments}

We conduct an experiment to test whether our model explains patterns of human reasoning about a variety of two-player grid-based games. We design $n=121$ grid game specifications (see examples in \fig{fig-novel-game-theory}A) which vary the game \textbf{environment} (including square boards varying from 3$\times$3 to 10$\times$10; rectangular boards such as 2$\times$5 and 4$\times$9; and boards specified over an infinite grid); the game \textbf{dynamics} (including games in which a given player opens by going twice); and the \textbf{player utility functions} (including games in which the first or second player has differing win conditions, like requiring the first player to make a larger row than the second one.) 
Table \ref{tab:stimuli} summarizes these game variants. 

\subsection{Human game evaluations and game construction}
We carry out a two-part study to evaluate how people reason about these novel games, and begin to to probe how humans create their own novel game variants to satisfy general game criteria (like constructing a reasonably \textit{fun} game).

\textbf{Participants} We recruit $484$ participants from Prolific \citep{palan2018prolific}. Each participant was randomly presented with $10$ games sampled from our stimuli, as well as regular Tic-Tac-Toe (won by making \textit{3 in a row on a $3\times3$ board}) to normalize game judgments. We collect approximately $20$ judgments per game stimuli for each game reasoning query. Participants were paid at a base rate of \$12.5/hr with an optional bonus up to \$15/hr; the full experiment approximately took $25$ minutes.

\textbf{Game outcome and game fun judgments} Subjects were randomly assigned to one of two game reasoning conditions. In the \textbf{game outcome evaluation} condition, subjects produced judgments on a continuous 0-100 probability scale to predict the likelihood of a first player win (\textit{if the game does not end in a draw, assuming both players play reasonably, how likely is it that the first player is going to win (not draw)?}) and a draw (\textit{assuming both players play reasonably, how likely is the game to end in a draw?}). In the \textbf{game fun} condition, subjects instead assessed the likelihood that the game is fun (\textit{how fun is this game}?) on a confidence scale spanning 0 (\textit{the least fun of this class of game}) to 100 (\textit{the most fun of this class of game}).

Participants produced judgments about each game based on a linguistic game specification. We additionally provided participants with an interactive \textit{scratchpad} board that they were told they could, but were not required, to use to inform their judgments; the scratchpad permitted automatically placing pieces of different colors (to simulate players) and could be cleared entirely to begin a new game. Participants were required to consider each game for at least $60$ seconds before producing game judgments; on average, participants took 87.2 ($\pm$ 2.5 STE) seconds per game in the game outcome condition and 83.4 ($\pm$ 2.41 STE) seconds per game in the fun rating condition.

\begin{figure*}[h!]
\begin{center}

\includegraphics[width=0.95\linewidth]{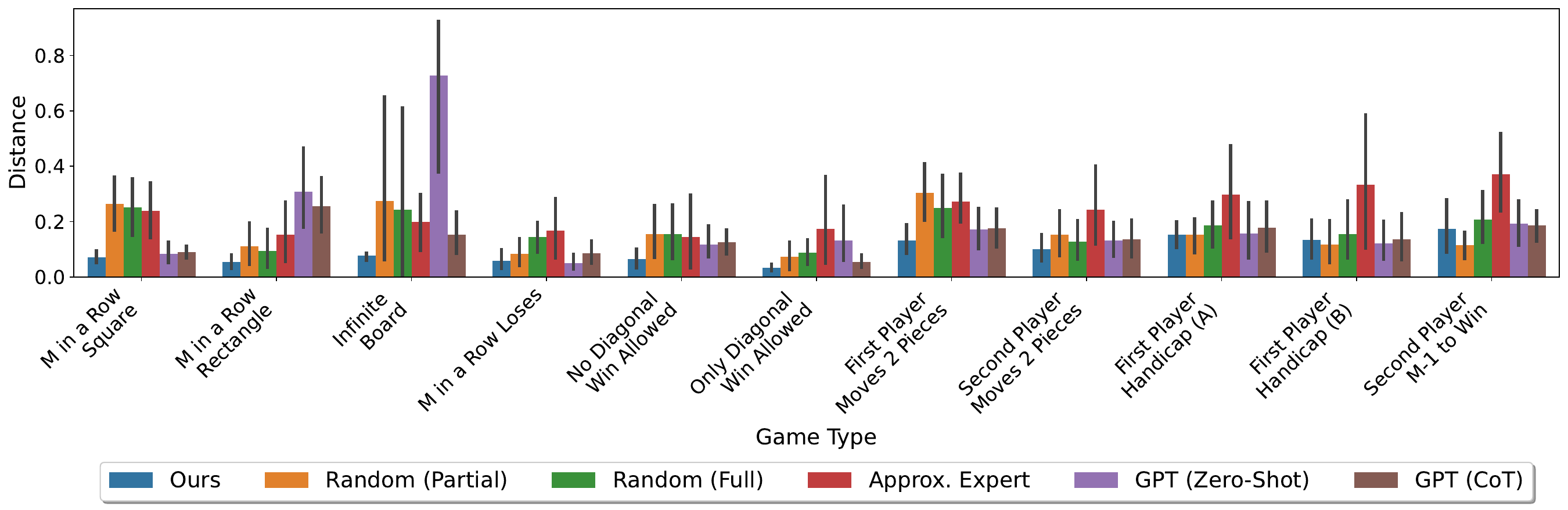}
\end{center}
\caption{Absolute difference between human- and model-predicted payoff, broken down by game category. Error bars depict 95\% CI over games within each game category (number of games in each game category is depicted in \ref{tab:stimuli}). }
\label{game-class-payoff}
\end{figure*}

\textbf{Novel game creation} After answering all game reasoning queries, we additionally asked participants to create a new grid-based game variant that they would \textit{find fun}. Participants wrote a linguistic game specification, describing the board size and win conditions. As in the game judgment queries, participants were again provided an optional scratchpad and required to spend $60$ seconds before submitting a response. The scratchpad enabled participants to try out the game they intend to create. After specifying a game, participants were asked to answer the same game reasoning query (either game outcomes or game fun) about their own game.

\subsection{Model game evaluations and alternative models} 
We estimate simulated game evaluation judgments under our model. We also implement a range of computational models designed to span both weaker estimates of game play (using more naive agent models, or without any explicit agent models) and \textit{more optimal} estimates of game play (using additional search and game simulations to terminal states).

\textbf{Ours (Partial game simulations, 1-step heuristic over actions):} Our model draws probabilistic game judgments via sample-based inference, where samples are partial game simulations rolled out under a subgoaling agent model that only estimates utilities over the immediate next actions according to equation \ref{eq:subgoal}. To match the sample size of the human experiment, we simulate $n=20$ judgments per game, where each judgment is estimated from $k= \ksamples$ samples.

\textbf{Partial game simulations, random actions:} This baseline estimates outcomes using partial game simulations, but substitutes our sub-goaling agent model with a more naive (``random'') agent model that selects actions from a uniform distribution over valid moves. We again simulate $n=20$ judgments per game and $k= \ksamples$ samples per judgment.

\textbf{Full game simulations, random actions:} This baselines infers outcomes from \textit{complete} game simulations to terminal states, but using the random agent model. We again simulate $n=20$ judgments per game and $k= \ksamples$ samples per judgment.

\textbf{Approximate expert model (full game simulations, heuristic look-ahead tree search)}: This baseline is designed to model a stronger, ``expert" player for each game, and infers outcomes by simulating play with a heuristic search algorithm that plans multiple steps ahead when choosing actions. This model is closely based on the model of human expertise for 4-in-a-row on a 4x9 board in \cite{van2023expertise}, which empirically estimates tree search depth from human players after hours of continuous gameplay experience\footnote{\cite{van2023expertise} estimates a probabilistic `stopping parameter' governing how many iterations over which they expand nodes in the search tree, where this parameter is fit to each individual expert. We run all simulations by expanding the search tree a fixed number ($k_{iterations}=636$) of iterations, where $k_{iterations}=636$ is the empirical mean value of this parameter estimated from the \cite{van2023expertise} data.}. To handle our diversity of games, we use an adapted, generalized heuristic that is directly comparable to novice model and can be evaluated on our broad range of win conditions, board sizes, and player sequences. Our approximate expert model probabilistically expands nodes in the search tree by repeatedly sampling actions from a softmax policy over its current action estimates, expanding unexplored states in the search tree, and backpropagating utilities estimated at future states. We include more details of our model in the appendix  (Algorithm \ref{alg:approximate-expert}). The expert model is significantly more computationally expensive than our main model. Due to computational constraints, we report expected payoffs based on just 50 \textit{total} simulations per game ($k=2.5$ samples per $n=20$ subjects). However, in general, this model has much less variance in payoff estimates even with small sample sizes, as utility estimates sharpen with further search.

\begin{figure*}[h!]
\begin{center}
\includegraphics[width=0.95\linewidth]{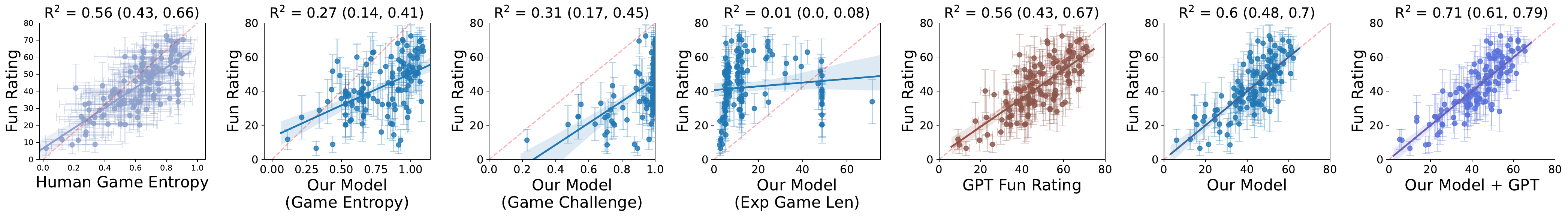}
\end{center}
\caption{Human game \textbf{fun ratings} correlate well with several distinct game features: entropy over game outcomes as predicted by participants themselves; outcome entropy predicted under our model; predicted advantage over a random agent given our model's game play; expected game length under our model; and LLM-based estimates given the game specification. Far right plots show correlations in a combined regression model fit to all features predicted under our model (entropy, advantage, and length); and additionally incorporating the LLM-based estimates. Error bars depict standard error. Numbers in parentheses indicate 95\% CI.}
\label{fun-rating-regression}
\end{figure*}

\textbf{Full game simulations with Monte Carlo Tree Search (MCTS)} This baseline infers outcomes using full game simulations played using an MCTS-based agent model \citep{genesereth2014general}. We use a standard, upper-confidence-bound MCTS implementation; pseudocode can be found in the appendix (Algorithm \ref{alg:mcts}). Unlike our main model or the approximate expert model, MCTS does not use game-specific heuristic features, and instead estimates intermediate utilities by expanding a search tree guided by repeated random rollouts to terminal states. MCTS algorithms are commonly used to approximate \textit{oracle} or optimal gameplay in arbitrary games \citep{genesereth2014general}, as they are empirically efficient but accurate estimators of optimal play for many games (they effectively explore the search tree to recover optimal utility estimates, and converge to optimal estimators in the limit). We use MCTS as an approximate `oracle' on our games and estimate actions using a large ($k=10,000$) number of tree expansions. As with the approximate expert model, due to computational costs, we estimate the expected payoffs using 50 simulations per game (2.5 simulations per subject) -- but again, this model has less variance in outcome estimates per game simulation, as it tends to approximate more optimal play.

\textbf{Large Language Model (LLM) game evaluations} We also implement two LLM baselines to evaluate game reasoning with no explicit agent model, but rather, models which can draw on background knowledge about classic grid games and variants (including extensive information, for instance, about the optimal solutions to games like Tic-Tac-Toe and Connect-4). We use GPT-4 \citep{openai2023gpt4} as our base LLM and implement two LLM-based game evaluation baselines: a \textit{zero-shot} baseline (which is directly queried with the linguistic game specification and game reasoning query), and a \textit{zero-shot chain-of-thought} (CoT) baseline \citep{kojima2022large, wei2022chain}, which is first prompted to produce additional reasoning in text before answering a query (a standard method used to account for the computational limits in directly producing answers to questions given the LLM architecture). In both baselines, we prompt the model with a lightly-modified variant of the full experiment instructions shown to human participants, and collect judgments on the same 0-100 scale for all queries.

\section{Results and Discussion}
Here we report our findings and discuss their interpretations.

\textbf{People's game outcome judgments are best predicted by partial, goal-directed game simulations.} 
Subjects provided judgments estimating both the likelihood of a draw, and the likelihood that the first player would win if the game did \textit{not} end in the draw, which collectively estimate the distribution and expected utility over all possible outcomes (draw, first player wins, second player wins) for any given game. \fig{overall-payoff} shows correlations between human estimations of game \textit{payoffs} (expected utility, where -1 is a loss and 1 is a win)---aggregated across all participants and game stimuli---and our model (\textit{Ours}, blue) as well as the alternative baselines. As shown in \fig{overall-payoff}, our model correlates well with human estimates ($R^2 = 0.86$).

Notably, our model is a much better fit to human predictions than both the more naive models (such as the \textit{random} partial simulation model, orange), \textit{and} the more optimal models (both the \textit{approximate expert}, red, and \textit{MCTS}, pink). In particular, the game outcome predictions under the approximate expert and the MCTS baselines reveal that under deeper search, there is much less uncertainty over game outcomes that does not match the gradedness and systematic deviations from optimality that humans demonstrate. A qualitative analysis of human scratchpad data supports the conclusion that while most subjects spontaneously do seem to simulate games, they draw game estimates from very few games (almost no participants reset and then reused the scratchpad more than $5$ times) and imperfect game simulations (see examples in \fig{results-hvm}).

\fig{game-class-payoff} shows mean absolute error between the predicted payoff under models and human judgments on games, further split among the subcategories of our novel game variants. As seen in \fig{game-class-payoff}, our model (blue) is close (and usually closest) to human judgments within arguably all but the last subcategories. We also find that the LLM-based models (both the \textit{zero-shot} and \textit{chain-of-thought} baselines) are in fact relatively close to human judgments for the games that are the most obviously related to Tic-Tac-Toe (the \textit{$M$ in a row} games on a square $N \times N$ board), but deviate much more from human judgments on seemingly minor variants---\textit{$M$ in a row} games on \textit{rectangular} boards or \textit{infinite} boards---possibly because these games are relatively out-of-distribution compared to the more familiar variants found in background linguistic data.

\textbf{People's game fun judgments can be predicted by their own and our model's game outcome estimates.} We evaluate several game-play-based features which we find each correlate with subject's \textbf{game fun} estimates: 1) an estimate of whether games are fair and balanced based on the \textit{entropy over game outcomes}---both predicted by humans themselves and predicted under our model; 2) an estimate of whether a game is reasonably challenging based on relative \textit{advantage}  between our model and a random agent (where expected utility is computed given random assignment of player order) using full game simulations; and 3) expected game \textit{length} under our model. Interestingly, we also find the LLM-based CoT estimates of game fun to be well-correlated with human judgments. \fig{fun-rating-regression} summarizes these individual features, as well as fits from regression models fit to our model-derived features (and including the additional LLM-based prediction). Better understanding what features of a game drive participants' funness judgments are ripe grounds for future work. 

\textbf{People spontaneously invent fair games when asked to invent fun games.} Finally, subject self-evaluations of their own \textit{invented} game variants (\fig{results-hvm}C) further supports the relationship between game outcomes and game fun: human evaluations suggest that people mostly generate games they believe are \textit{fair}. That is, participants endorse an unbiased expected payoff between players, when asked to generate games that they also self-evaluate as fun. 

\subsection{Conclusion and future directions}

Our findings demonstrate how people might make quick, probabilistic judgments about the expected value of complex novel problems---well before they learn narrower, task-specific strategies and representations that constitute genuine expertise---by integrating efficient, general-purpose computations for simulating multi-agent decision making and approximate probabilistic inferences. We see many avenues for future work. One is the question of how people grasp the constraints and goals of novel tasks, in order to reason about them. Here we assumed that people understood the linguistic descriptions of the games we presented them with, but this is a nontrivial cognitive process in its own right. Although large language models were limited in their ability to reason about \textit{new} games as people do, they have proven capable of semantic parsing and program synthesis, and we see opportunities for modeling language-based game understanding using LLMs together with probabilistic planning and decision models like our intuitive game theory models, in a neurosymbolic framework \citep{austin2021program, chen2021evaluating, collins2022structured, wong2023word}. 

Also, here we only modeled the earliest stages of playing and reasoning about novel games. It is plausible that people's skill in these games will rapidly increase as they play more. A deeper understanding of such fast concept and skill acquisition would be highly valuable \citep{lake2017building}. More broadly, the study of how people quickly learn to play novel games could contribute to contemporary work on resource rationality \citep{gershman2015computational, lieder2020resource, icard2023resource} and human efficient planning \citep{callaway2022rational, ho2022people}. It is an open question to what extent our current model can properly be called ``resource rational" and to what extent naive humans approximate answers to these kinds of intuitive questions using ``resource rational" representations. Overall, we believe general principles of efficient use of cognitive resources can be further applied to studying games, novel problem solving, and beyond.

\section{Acknowledgments}

We thank Graham Todd, Joy Hsu, Guy Davidson, Laura Schulz, Yuka Machino, Kartik Chandra, Tom Griffiths, and Ionatan Kuperwajs for helpful conversations that informed this work. CEZ, LW, and JBT acknowledge support from AFOSR Grant \#FA9550-19-1-0269, the MIT-IBM Watson AI Lab, MIT Quest for Intelligence, ONR Science of AI, and the Simons Center for the Social Brain. KMC acknowledges support from the Marshall Commission and the Cambridge Trust. AW acknowledges support from a Turing AI Fellowship under grant EP/V025279/1, The Alan Turing Institute, and the Leverhulme Trust via CFI.

\bibliographystyle{apacite}

\setlength{\bibleftmargin}{.125in}
\setlength{\bibindent}{-\bibleftmargin}

\bibliography{references}

\clearpage 
\newpage 

\section*{Appendix: Additional modeling details}

\subsection*{Approximate Expert model details}
Here we detail pseudocode for the approximate expert model (designed to generalize the 4-in-a-row model of human gameplay from \cite{van2023expertise}). Its algorithm repeats 3 sub-procedures, SelectNode, ExpandNode, and Backpropagate, within the main procedure MakeMove, which ultimately makes a move by sampling from a softmax distribution. 
\begin{algorithm}[h!]
\caption{Approximate Expert: MakeMove}\label{alg:approximate-expert}
\begin{algorithmic}
\State $root \gets \Node(s)$
\For{$i=1$ to $num\_steps$}
    \State $n\gets\SelectNode(root)$
    \State $\ExpandNode(n)$
    \State $\Backpropagate(n, root)$
\EndFor
\State \Return $\textrm{Sample} (\textrm{softmax} (\{c: c.val\textrm{ for }c\in root.children\})$
\end{algorithmic}
\end{algorithm}

The function Node is the constructor for the nodes of the tree search. In our simulations, we repeat the procedure in MakeMove for $num\_steps=636$ iterations, drawing on results from an empirical parameter fit in \cite{van2023expertise} which found that expert human players (playing a specific 4-in-a-row variant) were well modeled by a heuristic search procedure whose number of iterations is geometrically distributed with a mean stopping parameter of 1/636. Our model does not use a stochastic stopping parameter and just runs deterministically to $num\_steps=636$ iterations for all simulations.

\begin{algorithm}[h!]
\caption{Approximate Expert: $\SelectNode(n)$}\label{alg:ae-select-node}
\begin{algorithmic}
\State $root\gets n$
\While{$n.children\neq\emptyset$}
    \If{$n.player\_type=root.player\_type$}
        \State $n=\argmax_{c\in n.children} c.val$
    \Else
        \State $n=\argmin_{c\in n.children} c.val$
    \EndIf
\EndWhile
\State \Return $n$
\end{algorithmic}
\end{algorithm}
Here \textit{player\_type} $\in \{X, O\}$.

\begin{algorithm}[h!]
\caption{Approximate Expert: $\ExpandNode(n)$}\label{alg:ae-expand-node}
\begin{algorithmic}
\State $s\gets n.state$
\ForAll{$m\in \LegalMoves(s)$}
    \State $b\gets \PlacePiece(s,m,n.player\_type)$
    \State $i\gets n.move\_number$
    \State $n.children.append(\Node(b,i+1, V^\AE(b))$
\EndFor
\end{algorithmic}
\end{algorithm}

Here \textit{move\_number} is the index of the current move over the entire game (to account for variants which specify that a given player goes twice as their opening move).

Within the procedure ExpandNode, the function $V^\AE$ is a modified version of our heuristic function $V$ defined in equation \ref{eq:subgoal} that depends only on the board state $b$ and is symmetric for both players. To set a frame of reference, suppose player $X$ is evaluating the utility of board state $b$. The heuristic $V^\AE(b)$ is evaluated as $V_X^\AE(b)-V_O^\AE(b)$ where 
$$
    V^\AE_a(b)=\sum_{p\in a\textrm{ pieces}}V(p,b).
$$
for $a\in \{X,O\}$. The summation is over all pieces $p$ on the board that $a$ has placed. When $O$ acts, the heuristic is symmetrically $V^\AE(b)=V^\AE_O(b)-V^\AE_X(b)$. We make this modification so that $V^\AE$ is comparable across different branches of the search tree and so that we can efficiently alternate between the two players in the tree. 

\begin{algorithm}[h!]
\caption{Approximate Expert: $\Backpropagate(n, root)$}\label{alg:ae-backpropagate}
\begin{algorithmic}
\If{$n.player\_type=root.player\_type$}
    \State $n.val \gets \max_{c\in n.children}c.val$
\Else
    \State $n.val\gets\min_{c\in n.children}c.val$
\EndIf
\end{algorithmic}
\end{algorithm}

\subsection*{MCTS model details}
Similar to the approximate expert model, MCTS repeats 4 sub-procedures---SelectNode, ExpandNode, DepthCharge and Backpropagate, but the SelectNode, ExpandNode, and Backpropagate sub-procedures are different from their corresponding counterparts in the approximate expert model. We run MCTS for 10,000 steps, which we find balances a close approximation of the optimal for many games while keeping the runtime of any single game trial to less than 2 days.

\begin{algorithm}[h!]
\caption{MCTS: MakeMove}\label{alg:mcts}
\begin{algorithmic}
\State $root \gets \Node(s)$
\For{$i=1$ to $num\_steps$}
    \State $n\gets\SelectNode(root)$
    \State $\ExpandNode(n)$
    \State $game\_outcome = \DepthCharge(n, root)$
    \State $\Backpropagate(n, root, game\_outcome)$
\EndFor
\State \Return $\argmax_{c\in root.children}c.val$
\end{algorithmic}
\end{algorithm}

\begin{algorithm}[h!]
\caption{MCTS: $\SelectNode(n)$}\label{alg:mcts-select-node}
\begin{algorithmic}
\While{$n.children\neq\emptyset$}
    \State $n=\argmax_{c\in n.children} c.UCB$
\EndWhile
\State \Return $n$
\end{algorithmic}
\end{algorithm}
To balance exploring and exploiting in the MCTS SelectNode sub-procedure, we use a standard upper confidence bound (UCB) \citep{browne2012survey}. 
\begin{align*}
c.UCB = &\argmax_{c\in root.children}\big(\frac{c.val}{c.visits} \\ &+\sqrt{2\frac{\log(c.parent.visits)}{c.visits}}\big). 
\end{align*}
\begin{algorithm}[h!]
\caption{MCTS: $\ExpandNode(n, root)$}\label{alg:mcts-expand-node}
\begin{algorithmic}
\State $s\gets n.state$
\ForAll{$m\in \LegalMoves(s)$}
    \State $b\gets \PlacePiece(s,m,n.player\_type)$
    \State $i\gets n.move\_number$
    \State $n.children.append(\Node(b,i+1))$
\EndFor
\end{algorithmic}
\end{algorithm}

\begin{algorithm}[h!]
\caption{MCTS: $\Backpropagate(n, root, game\_outcome)$}\label{alg:mcts-backpropagate}
\begin{algorithmic}
\State $n.visits \gets n.visits+1$
\If{n.parent}
    \If{n.parent.player\_type == root.player\_type}
        \State $n.val \gets n.val+game\_outcome$
    \Else
        \State $n.val \gets n.val+(1-game\_outcome)$
    \EndIf
    \State $\Backpropagate(n.parent, root, game\_outcome)$
\EndIf
\end{algorithmic}
\end{algorithm}

\begin{algorithm}[h!]
\caption{MCTS: $\DepthCharge(n)$}\label{alg:mcts-depthcharge}
\begin{algorithmic}
\State $s\gets n.state$
\State $i\gets n.move\_number$
\While{not $\HasTerminated(s)$}
    \State $player\_type\gets n.player\_sequence[i]$
    \State $m\gets \RandMove(s)$
    \State $i\gets i+1$
    \State $s\gets\PlacePiece(s,m,player\_type)$
\EndWhile
\If{$\IsWin(s, root.player\_type;game\_rules)$}
    \State \Return 1
\ElsIf{$\IsDraw(s; game\_rules)$}
    \State \Return 1/2
\Else
    \State \Return 0
\EndIf
\end{algorithmic}
\end{algorithm}

Finally, as noted above, our implementations of the approximate expert and MCTS algorithms always parameterize simulation with a formal specification of the rules of any given game (e.g., restrictions on whether a given player can win only in certain directions on the board). We implement $V$ and DepthCharge as described in the main text so that intermediate value functions and win conditions are calculated with respect to the specific game dynamics and win conditions for each game variant. \textit{MakeMove} similarly takes in the current \textit{move\_number} move index to account for games in which player order is not strictly alternating, such as games in which the \textit{first player goes twice} as their opening move. 

\end{document}